\documentclass[journal=jacsat,manuscript=article]{achemso}

\usepackage{chemformula} 
\usepackage[T1]{fontenc} 

\usepackage{graphicx}
\usepackage{dcolumn}
\usepackage{color}
\usepackage{bm}
\usepackage[utf8]{inputenc}
\usepackage[pdftex,colorlinks]{hyperref}
\hypersetup{ 
urlcolor=blue,
linkcolor=blue,
citecolor=blue,
pagebackref=true
}

\title{Germanium quantum well Josephson field effect transistors and interferometers} 
\author{F. Vigneau}
\author{R. Mizokuchi}
\author{D.  Colao Zanuz}
\affiliation{Univ. Grenoble Alpes, CEA, INAC-Pheliqs, 38000 Grenoble, France}
\author{X. Huang}
\affiliation{Department of Physics and Astronomy, University of Pittsburgh, Pittsburgh, PA 15260}
\author{S. Tan}
\affiliation{Department of Electrical and Computer Engineering and Petersen Institute of NanoScience and Engineering, University of Pittsburgh, Pittsburgh, PA 15260}
\author{R. Maurand}
\affiliation{Univ. Grenoble Alpes, CEA, INAC-Pheliqs, 38000 Grenoble, France}
\author{S. Frolov}
\affiliation{Department of Physics and Astronomy, University of Pittsburgh, Pittsburgh, PA 15260}
\author{A. Sammak}
\affiliation{QuTech and Kavli Institute of Nanoscience, Delft University of Technology Lorentzweg 1, 2628 CJ Delft, Netherlands}
\affiliation{QuTech and TNO, Stieltjesweg 1, 2628 CK Delft, The Netherlands}
\author{G. Scappucci}
\affiliation{QuTech and Kavli Institute of Nanoscience, Delft University of Technology Lorentzweg 1, 2628 CJ Delft, Netherlands}
\author{F. Lefloch}
\author{S. De Franceschi}
\affiliation{Univ. Grenoble Alpes, CEA, INAC-Pheliqs, 38000 Grenoble, France}

\email{silvano.defranceschi@cea.fr}

\begin{document}

\date{\today}

\begin{abstract}

Hybrid superconductor-semiconductor structures attract increasing attention owing to a variety of potential applications in quantum computing devices. They can serve to the realization of topological superconducting systems, as well as gate-tunable superconducting quantum bits. Here we combine a SiGe/Ge/SiGe quantum-well heterostructure hosting high-mobility two-dimensional holes and aluminum superconducting leads to realize prototypical hybrid devices, such as Josephson field-effect transistors (JoFETs) and superconducting quantum interference devices (SQUIDs). We observe gate-controlled supercurrent transport with Ge channels as long as one micrometer and estimate the induced superconducting gap from tunnel spectroscopy measurements in superconducting point-contact devices. Transmission electron microscopy reveals the diffusion of Ge into the aluminum contacts, whereas no aluminum is detected in the Ge channel. 

\end{abstract}

\maketitle

Modern quantum nanoelectronics takes increasing advantage of newly synthesized hybrid superconductor-semiconductor (S-Sm) interfaces \cite{Silvano2010}. One of the main motivations is the search for Majorana zero modes that are predicted to appear in a topological superconductor \cite{kitaev2001, oreg2010, stanescu2010}. A Josephson field effect transistor (JoFET) is one of the basic devices. It consists of a gate-tunable semiconductor channel allowing Cooper-pair exchange between two superconducting contacts mediated by the superconducting proximity effect. \cite{clark1980}. Gate control on the Josephson coupling has eventually led to the realization of electrically tunable transmon quantum bits, now often referred to as gatemons \cite{larsen2015, Lange2015, Casparis2018}.

 Many of the reported experimental realizations of hybrid S-Sm devices rely on bottom-up fabrication starting from semiconductor nanowires or carbon nanotubes \cite{chang2015, mourik2012,zhang2017,gul2017,de2018, xiang2006, su2016, jarillo2006}. Recently, new hybrid S-Sm devices were demonstrated using top-down fabrication processes based on two-dimensional systems made of graphene \cite{heersche2007}, InAs \cite{kjaergaard2016,Shabani2016}, GaAs \cite{wan2015}, InGaAs \cite{delfanazari2018} or Ge/SiGe \cite{hendrickx2018, hendrickx2018ballistic}. 

Top-down nanoscale devices offer significant advantages in terms of complexity and scalability. Those based on p-type SiGe heterostructures are readily compatible with silicon technology \cite{pillarisetty2011}, and, thanks to their intrinsically strong spin-orbit coupling, they are an attractive candidate for the development of topological superconducting systems \cite{hendrickx2018, morrison2016, Watzinger2016, Ares2013, Kloeffel2011, morrison2014, Zarassi2017, mizokuchi2017, mizokuchi2018}. 

In this work, we present proof-of-concept S-Sm devices in which the semiconducting element consists of an undoped SiGe heterostucture embedding a strained Ge quantum-well (QW). A high-mobility two-dimensional hole gas (2DHG) is electrostatically accumulated in the QW by means of a surface gate electrode. (Hole mobilities as high as $5\times 10^{5}$ cm$^2$/Vs were reported for similar heterostructures \cite{sammak2018, hendrickx2018, myronov2014, gul2017}.) The superconducting proximity effect induces gate-tunable superconductivity in the 2DHG enabling JoFET operation. This functionality is exploited for the realization of gate-controlled superconducting quantum interference devices (SQUIDs). Finally, We present tunnel-spectroscopy measurements of the induced superconducting gap, as well as a microscopic inspection of the S-Sm interface by cross-sectional transmission-electron-microscopy.

\begin{figure}[ht]{\centerline{\includegraphics[width=1\columnwidth,clip,angle=0]{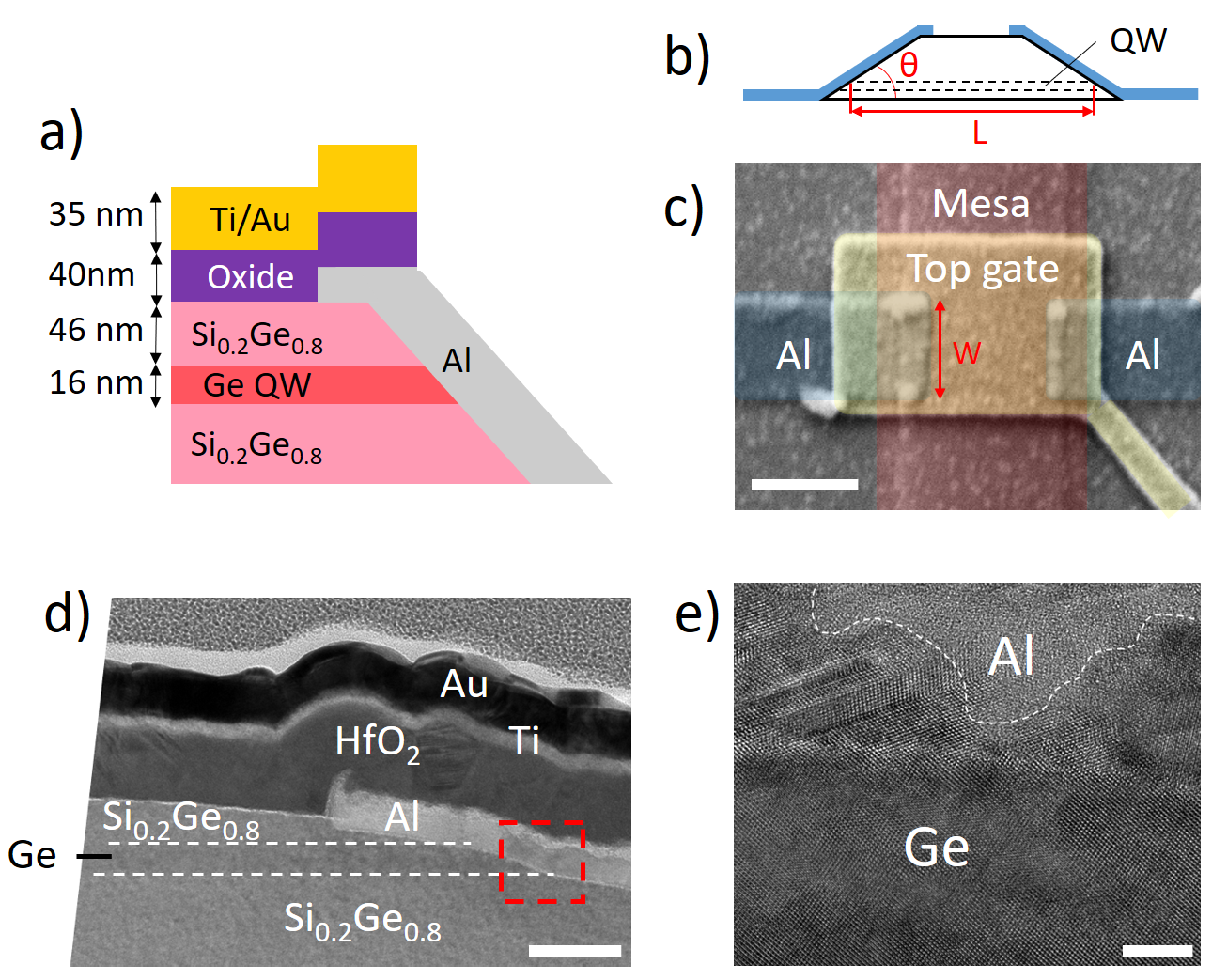}}}
\caption {a) and b) Schematic cross section of the quantum well and device structure.
c) False color SEM picture of a device consisting of a top gate covering two Al contacts and a mesa. The scale bar is 500\,nm.
d) TEM cross-sectional image of a single-junction device cut at the level of the Al - Ge QW junction. The scale bar is 50\,nm.
e) High resolution TEM image of the Ge - Al interface in the area inside the red dash square in (a). The scale bar is 5\,nm.
}\label{FigTEM}
\end{figure}

All devices have Al superconducting leads deposited on the sidewalls of a wet-etched mesa structure. The etching process yields a sidewall slope of $\theta \approx 10^\circ$ enabling direct side contacts to the QW [Fig. \ref{FigTEM}(a,b)]. The channel length, $L$, defined by the distance between the two Al/Ge contacts, ranges between 0.6 and 1 $\mu$m. The surface gate electrodes are made of Ti/Au and they are fabricated after the atomic layer deposition (ALD) of an insulating oxide layer [Fig.\ref{FigTEM}(c)]. We distinguish between two types of top gates: those aimed at inducing hole accumulation to create a conducting hole channel; and those aimed at causing local charge depletion and a resulting channel constriction.

We investigate the interface between the Al contacts and the Ge QW with TEM analysis. First, we observe direct contact between the Al and the QW [Fig. \ref{FigTEM} (d)]. Second, diffusion of Ge atoms into the Al layer is found at the Al/QW interface, seen as darker contrast in Fig. \ref{FigTEM} (d,e) on the Al side opposite of the QW and confirmed by EDS analysis (see Supporting Information). We speculate that this occurs during the gate oxide deposition by ALD, when the samples are heated at 250\,$^{\circ}$C for 2.5 hours. Under similar conditions, the diffusion of Al (or Pt) atoms through a SiGe barrier into the Ge QW has been observed and used to produce low resistivity electrical contacts from the top \cite{kral2015, hendrickx2018, de2018, xiang2006, su2016, sammak2018}. In general, interdiffusion at the S-Sm interface can lead to high critical supercurrents due to reduced interface resistance. However, we do not find evidence of Al diffusion down through SiGe barriers or from the side directly into the Ge QW in these devices.

In order to demonstrate JoFET functionality, we perform two-probe transport measurements on a single junction device at 15\,mK using a standard lock-in technique. The sample is in series with low-temperature RC filters and high-frequency filters. All transport data shown in this work are corrected to remove the contribution from the total series resistance of the measurement circuit (about 45\,k$\Omega$). 

\begin{figure}[ht]{\centerline{\includegraphics[width=1\columnwidth,clip,angle=0]{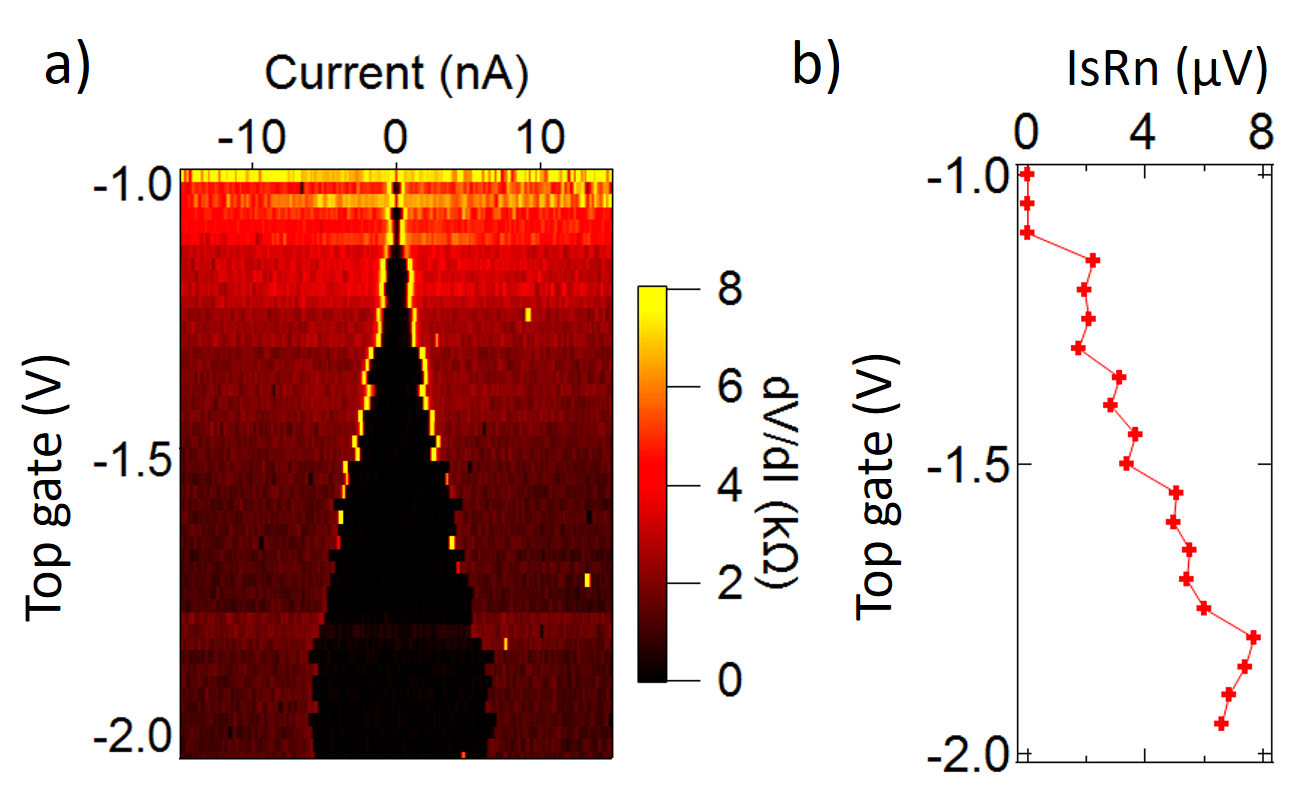}}}
\caption {a) Differential resistance of the single junction device versus top gate voltage and current. 
b) $I_sR_n$ product versus gate voltage from the results presented in (a). 
}\label{FigJoFET}
\end{figure}

Figure \ref{FigJoFET}(a) shows a measurement of differential resistance ($dV/dI$) as a function of bias current $I$ and top gate voltage $V_g$ in a single-junction device similar to the one shown in Fig. \ref{FigTEM}(c) with $L$ = 1\,$\mu$m. A negative top gate voltage $V_g$ < -0.9\,V induces the accumulation of holes in the strained Ge channel, as revealed by the onset of channel conduction. Note that the gate-induced electric field is partially screened at the contacts due to the Al source and drain electrodes partially overlapping the mesa. Nevertheless, this does not appear to prevent the injection of holes from the Al electrodes. By making $V_g$ more and more negative, the zero-bias device resistance drops and eventually vanishes (this enables an accurate measurement of the circuit series resistance). Above $I_s$ the device resistance takes a finite value $R_n$ of the order of a few k$\Omega$.

As $V_g$ is tuned towards more negative values, $I_s$ increases while $R_n$ decreases. The observed gate-voltage dependence of $I_s$ demonstrates JoFET operation. 
$I_s$ and $dV/dI$ are roughly symmetric around zero bias, which is characteristic of an overdamped regime. The product I$_s$R$_n$ reaches the highest value of about 8\,$\mu V$ for the most negative gate voltages [Fig. \ref{FigJoFET}(b)]. The observed $I_sR_n$ is similar to the values recently reported for Ge 2DHG devices \cite{hendrickx2018} but small in comparison to those reported for Ge-Si core-shell nanowires \cite{xiang2006, abay2014, ridderbos2018, su2016}. The discrepancy may be due to significantly longer channel length in 2DHG devices.

\begin{figure}[ht]{\centerline{\includegraphics[width=1\columnwidth,clip,angle=0]{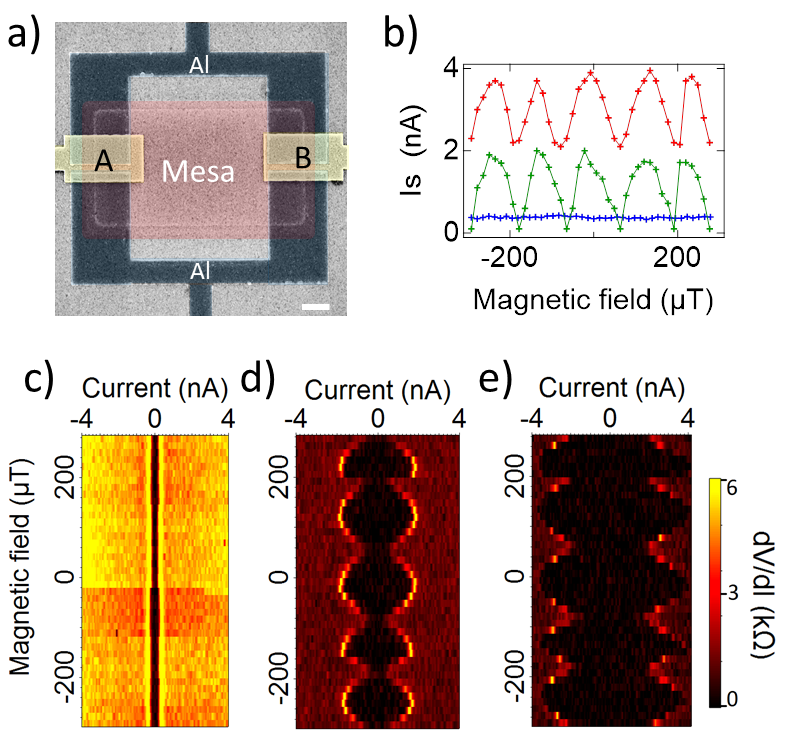}}}
\caption {a) False color SEM picture of a SQUID device. The scale bar is 1\,$\mu$m. 
b) Switching current versus magnetic field in three different regimes obtained from differential resistance measurements shown in (c) blue, (d) green and (e) red. 
c)-e) Differential resistance versus current and magnetic field measured for gate A = -980\,mV and different settings of gate B: -300\,mV (c), -500\,mV (d), -600\,mV (e).
}\label{FigSQUID}
\end{figure}

Next, we present the realization of SQUIDs consisting of two independently controlled JoFETs position in the two arms of a superconducting aluminum ring with an inner (outer) surface area of about 12 (27.5) $\mu$m$^2$ [Fig. \ref{FigSQUID} (a)]. The JoFET layout in this SQUID geometry is noticeably different from the one discussed above [Fig.\ref{FigTEM}(c)]. In fact, both superconducting contacts and the accumulated hole channel in between now lie on the same edge of the mesa. This geometry is less susceptible to the screening of the gate-induced electric field by the aluminum contact electrodes. On the other hand, the hole mobility in the Ge channel is most likely lowered by the proximity to mesa edge.

To investigate SQUID operation, we measure dV/dI in a four-probe configuration and apply a magnetic field, $B$, perpendicular to the device plane. We begin by characterizing one Josephson junction at a time. This is straightforward since the junctions are off (i.e. fully insulating) when no voltage is applied to the respective gates. We find that both junctions have $I_sR_n$ values consistent with those obtained for the device geometry of Fig.  \ref{FigJoFET}. The results are presented in Supplementary Materials. 

When both junctions are simultaneously on (i.e. in the superconducting regime) we can observe periodic SQUID oscillations in the switching current as a function of the out-of-plane magnetic field. Representative results are given in Figs. \ref{FigSQUID}(c-e), showing color maps of $dV/dI$ as a function of bias current and $B$. The three data sets are obtained with a fixed voltage on gate A, turning the left junction on, and three different voltages on gate B, resulting in correspondingly different values of the switching current in the right junction.  

We note that when the right junction is off, the supercurrent is carried uniquely by the left junction, and $I_s$ shows no B-induced oscillations [Fig. \ref{FigSQUID} (c)].

When both junctions are turned on, $I_s$ exhibits pronounced periodic oscillations [Fig. \ref{FigSQUID} (d)]. The modulation period $\Delta B_0 \approx 120 \, \mu$T corresponds to a magnetic flux quantum $h/2e$ threading an area of 17.2 $\mu m^2$, which is consistent with the size of Al ring. At the minima, $I_s$ is close to zero suggesting that  for this gate voltages the two junctions have approximately the same critical current \cite{tinkham2004}. 

After tuning the voltage on gate B to further negative values, the right junction becomes dominant and the SQUID is no longer balanced, as shown in Fig. \ref{FigSQUID} (e) . In this configuration, the average value of $I_s$ increases while the oscillation amplitude, which is determined by the smaller critical current of junction A, remains the same as in panel (b) \cite{van2006}. As a consequence, $I_s$ no longer vanish at the minima.

\begin{figure}[ht]{\centerline{\includegraphics[width=1\columnwidth,clip,angle=0]{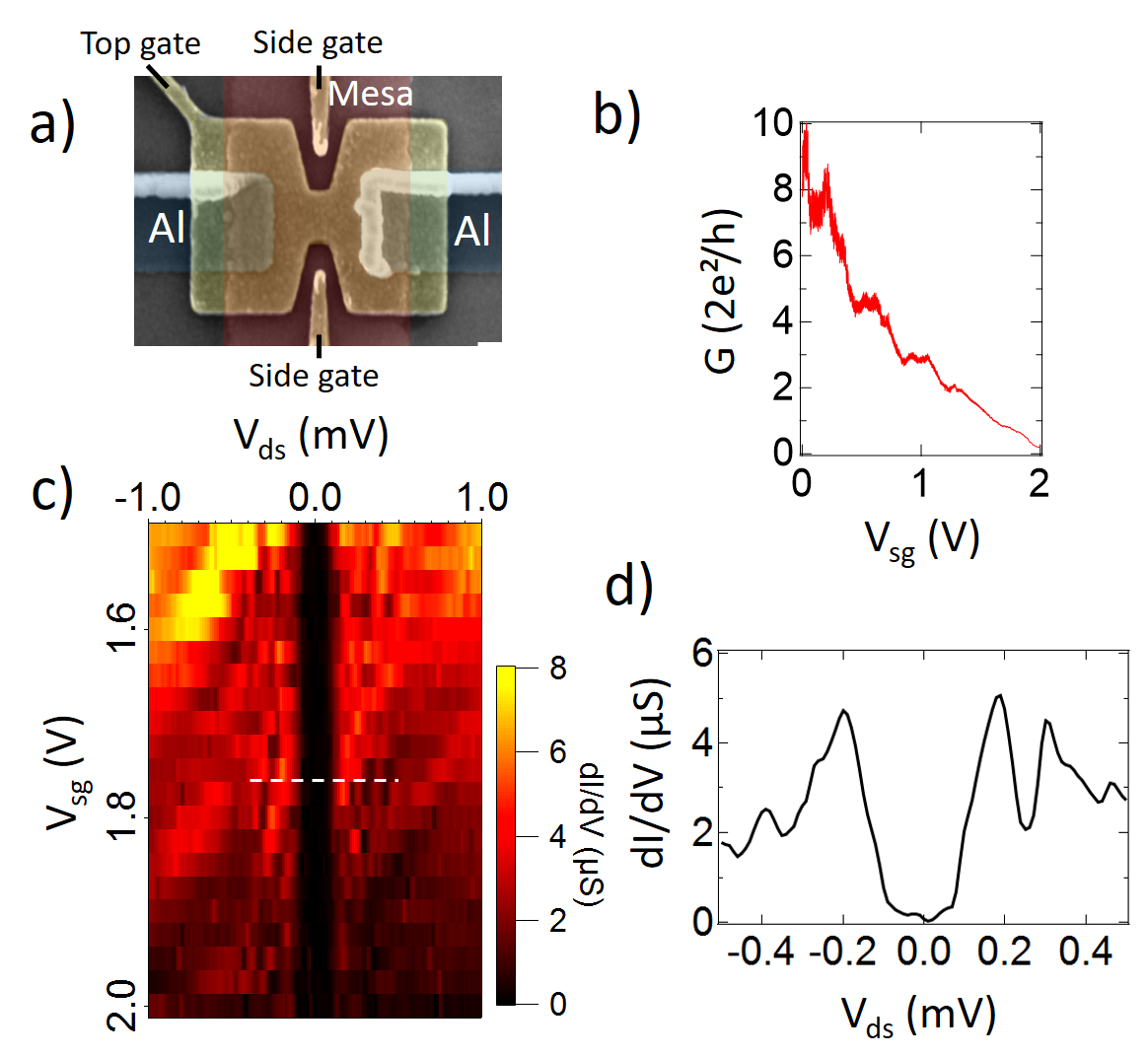}}}
\caption {a) False color SEM picture from side gate device. The scale bar represent 200\,nm. 
b) Differential conductance versus side gate voltage for top gate voltage -1.45\,V measured at 2\,mV of bias voltage.
c) Differential conductance versus side gate voltage and bias voltage measured at top gate voltage -1.41\,V
d) Line cut of figure (c) at side gate voltage indicated by the white dash line 1.7\,V)
}\label{FigQPC}
\end{figure}

In order to probe the superconducting gap $\Delta^*$, induced in the Ge 2DHG we use a JoFET with a constriction in the accumulation top gate and additional side gates designed to create a quantum point contact with tunable transmission [Fig. \ref{FigQPC}(a)]. In an earlier work by some of us \cite{mizokuchi2018}, the same device geometry was used to demonstrate conductance quantization in non-superconducting devices. In the regime of full depletion, the potential barrier at the constriction can be used to perform a tunnel spectroscopy of the local density of states.

As before, the top gate is negatively biased in order to accumulate holes and make the channel conducting. Then we apply a positive voltage V$_{sg}$ to both side gates simultaneously. The conductance pinch-off characteristic exhibits plateau-like structures reminiscent of conductance quantization [Fig. \ref{FigQPC} (b)]. However, these features are not as clear as in Ref. \cite{mizokuchi2018}, where a somewhat different SiGe heterostructure was used.

We are interested in the tunneling regime at high positive V$_{sg}$. Figure \ref{FigQPC} (c) shows a color map of the differential conductance, $dI/dV$, as a function of source-drain bias voltage, $V_{sd}$, and V$_{sg}$. We observe a region of suppressed $dI/dV$ around $V_{sd}=0$ [Fig. \ref{FigQPC} (d)], which is due to the presence of an induced superconducting gap centered around the Fermi energy. Based on the expected behavior of superconductor-insulator-superconductor tunnel junctions, the region of suppressed $dI/dV$ should extend between $V_{sd}=-2\Delta^*/e$ and $V_{sd}=2\Delta^*/e$, assuming the same $\Delta^*$ on both sides of the tunnel point contact. According to this hypothesis, we estimate $\Delta^*= 0.10$ meV. 
Gate-dependent peak/dip structures can been seen outside the gap region. We attribute them to mesoscopic resonances associated with quasi-bound states in the channel on either sides of the gate-defined middle barrier.

The estimated $\Delta^*$ is roughly half of the superconducting gap in bulk Al. 
We should like to note that we cannot exclude the unlikely, yet possible, presence of a strong asymmetry between the two sides of the tunnel point contact. 
For instance, a significantly less transparent contact on the left side would lead to a correspondingly weaker proximity effect and a "soft" induced gap. As a result, the suppressed conductance in Fig. \ref{FigQPC} (c) would rather be associated with the superconducting gap induced on the right side only, which would imply $\Delta^* = 0.20$ meV.

In conclusion, this work provides ample evidence of gate-tunable induced superconductivity in a high-mobility 2DHG confined to a Ge QW. JoFET functionality in single junctions is corroborated by the observation of $h/2e$-periodic oscillations in the switching current of SQUID-type devices, which can find application in phase sensitive experiments \cite{van2006, szombati2016, fornieri2018}. Our device processing approach permits a direct contact between the Al-based superconducting electrodes and the Ge QW. In spite of some unintentional out-diffusion of Ge into Al, the contact transparency is high, as denoted by the observation of an induced gap comparable to the one of bulk aluminum. The size of the contacts is in 100 nm range and in could be further reduced enabling ample versatility in device design. Finally, the developed contact scheme can be readily applied to other types of superconducting materials.

\section{Fabrication method}

High mobility SiGe/Ge/SiGe heterostructures were grown by reduced pressure chemical vapor deposition (see Figure \ref{FigTEM}(a) for a cross-sectional diagram). Details on growth process can be found in Ref. \cite{sammak2018}.
Mesa structures have a typical of depth 80\,nm and lateral sizes in the widths $\mu$m range. They are defined by e-beam lithography and chemical wet etching in a solution of H$_2$O:HF(10\%):HNO$_3$(69,5\%) 2:1:2.6 \cite{xue2011}. Contacts are defined by e-beam lithography and e-beam evaporation of 20\,nm aluminum \cite{wan2015}. Argon plasma etching is used to remove the native oxide prior to metal deposition. Then an dielectric layer of 30-40\,nm HfO$_2$ (Al$_2$O$_3$ for the side gate device) is deposited by ALD at a temperature of 250\,$^{\circ}$C. Finally 10/25\,nm Ti/Au top gate is patterned by e-beam lithography followed by metal e-beal evaporation and lift-off.

\section{Acknowledgement}
We acknowledge financial support from the Agence Nationale de la Recherche, through the TOPONANO and PIRE-HYBRID projects. Work is further supported by NSF PIRE-1743717, Grenoble Nanoscience Foundation, and the CEA program DRF impulsion Super-G. 

\providecommand{\latin}[1]{#1}
\makeatletter
\providecommand{\doi}
  {\begingroup\let\do\@makeother\dospecials
  \catcode`\{=1 \catcode`\}=2 \doi@aux}
\providecommand{\doi@aux}[1]{\endgroup\texttt{#1}}
\makeatother
\providecommand*\mcitethebibliography{\thebibliography}
\csname @ifundefined\endcsname{endmcitethebibliography}
  {\let\endmcitethebibliography\endthebibliography}{}

\end{document}